\documentclass[preprint,aps]{revtex4}  
\usepackage{graphicx}

\begin{document}

\title{Sub shot noise phase quadrature measurement of intense light beams}

\author{O.~Gl\"ockl, U.~L.~Andersen, S.~Lorenz, Ch.~Silberhorn*, N.~Korolkova,
G. Leuchs} \address{Institut f\"ur Optik, Information und Photonik,
Max--Planck Forschungsgruppe, Universit\"at Erlangen--N\"urnberg,
Staudtstr.~7/B2, 91058 Erlangen, Germany}

\begin{abstract}
We present a setup to perform sub shot noise measurements of the phase
quadrature  for intense pulsed light without the use of a separate local
oscillator. A Mach--Zehnder interferometer with an unbalanced arm length
is used to detect the fluctuations of the phase quadrature at a single side band
frequency. Using this setup, the non--separability of a pair of
quadrature entangled beams is demonstrated experimentally.\\
OCIS codes: 270.0270, 270.2500, 270.5570

\end{abstract}


\maketitle 

For many applications in quantum communication with continuous variables, such
as quantum teleportation\cite{FUR98}, entanglement swapping\cite{GLOE03}
and quantum cryptography\cite{GRO03}, it is required that one measures the
amplitude and the phase quadrature of the electromagnetic field. A homodyne
detector\cite{YUE83} is usually applied to perform phase sensitive measurements
by the interference of the signal beam and probing of the sidebands with a local
oscillator, which is required to be much brighter than the signal beam. However,
for intense signal beams, this requirement gives rise to technical difficulties
because the high intensities may saturate the detectors. Phase measurements can
also be achieved by rotating the bright carrier (internal local oscillator) with
respect to the sidebands. Such frequency dependent phase shifts could be
accomplished by reflecting the light of a cavity as a result of multiple beam
interference. This technique was used in early quantum optical
experiments\cite{SHE86,BAC88}. However, for ultrashort light pulse trains the
requirements on the dispersion properties of the resonator are quite demanding.
In this Letter, we present an alternative approach in which the fluctuations of
the phase quadrature are measured at a certain sideband frequency without the
use of any local oscillator or resonator. An interferometric setup reminiscent
of that used by Inoue and Yamamoto to determine the longitudinal mode partition
noise\cite{INO97} is shown to be capable of performing quantum--optical
measurements of the phase quadrature below the shot noise level, by introducing
a phase shift between the carrier and the sidebands due to two beam
interference. Our setup allows for easy switching between the measurement of the
phase quadrature and the amplitude quadrature and was used to fully characterize
quadrature entanglement of a pair of intense pulsed beams.

\begin{figure}[h]
\includegraphics{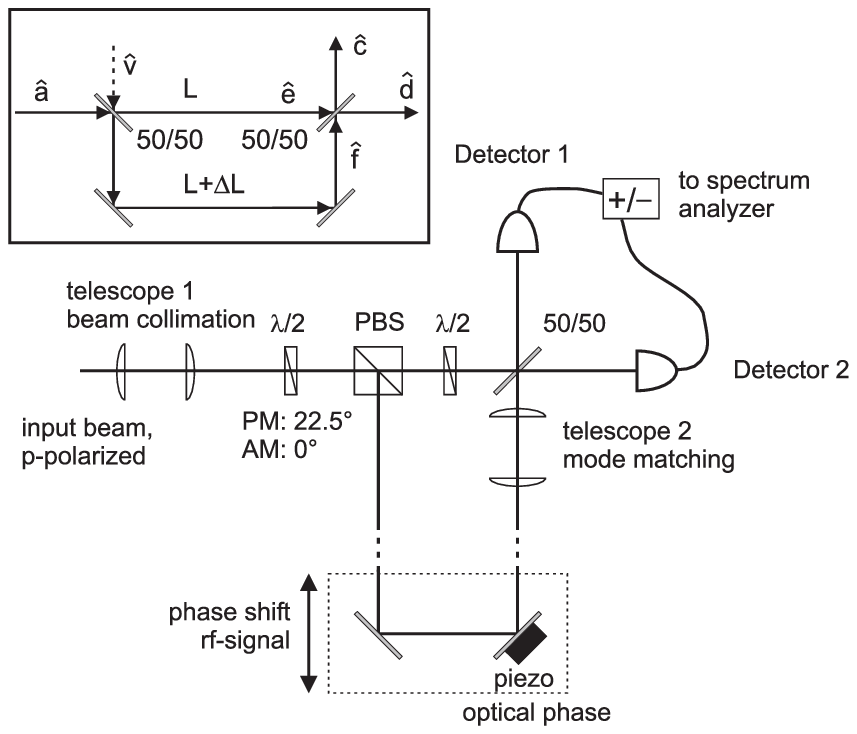}
\caption{\label{bild2}Experimental setup of the phase measuring interferometer.
The orientation of the first $\lambda/2$--plate determines the type of
measurement: amplitude measurement (AM) or phase measurement (PM). PBS,
polarizing beam splitter; 50/50, beam splitter} \end{figure}

The basic setup is a Mach--Zehnder interferometer (see the inset in Fig.
\ref{bild2}), with an arm length difference $\Delta L$ that introduces a
frequency dependent phase shift at a certain rf--sideband. For frequencies in
the MHz--regime, $\Delta L$ is much larger than the optical wavelength,
typically several meters. To describe the interferometer quantum theoretically
we express the input field mode through annihilation operator
$\hat{a}(t)=\alpha + \delta \hat{a}(t)$, where $\alpha$ is the classical
amplitude ($\alpha$ is assumed real) and where $\delta\hat{a}(t)$ contains
all classical and quantum mechanical fluctuations and having zero mean value.
This mode is split at a 50/50 beamsplitter and hence mixed with vacuum
$\delta\hat{v}(t)$. The two emerging modes denoted $\hat{e}(t)$ and $\hat{f}(t)$
are combined at a second 50/50 beamsplitter. The relative optical phase shift
between $\hat{e}(t)$ and $\hat{f}(t)$ is denoted $\varphi$, the relative delay
between the long and the short arms of the interferometer is given by
$\tau=\Delta L / c$. Output modes
$\hat{c}(t)=1/\sqrt{2}(\hat{e}(t)+\hat{f}(t))$ and
$\hat{d}(t)=1/\sqrt{2}(\hat{e}(t)-\hat{f}(t))$ are then given by
\begin{eqnarray} \hat{c}(t)= \frac{1}{2}\left(\alpha + \delta\hat{a}(t)
+\delta\hat{v}(t)+ {\rm e}^{i\varphi}\alpha+ {\rm e}^{i\varphi}\delta
\hat{a}(t-\tau)- {\rm e}^{i\varphi}\delta \hat{v}(t-\tau)\right),\\
\hat{d}(t)=
\frac{1}{2}\left(\alpha + \delta\hat{a}(t) +\delta\hat{v}(t)-{\rm
e}^{i\varphi}\alpha- {\rm e}^{i\varphi}\delta \hat{a}(t-\tau)+ {\rm
e}^{i\varphi}\delta \hat{v}(t-\tau)\right). \end{eqnarray}
For intense states of light, the interferometer
equations can be simplified by use of a linearization approach. Photon
numbers $\hat{n}_{\rm c}(t)=\hat{c}^{\dagger}\hat{c}$ and $\hat{n}_{\rm
d}(t)=\hat{d}^{\dagger}\hat{d}$ in the two output ports of the interferometer
are calculated by keeping the fluctuating contributions up to linear terms.
Evaluation of the sum $\hat{n}_{\rm c}(t)+\hat{n}_{\rm d}(t)$ and the difference
$\hat{n}_{\rm c}(t)-\hat{n}_{\rm d}(t)$ of the photocurrents yields
\begin{eqnarray}
\hat{n}_{\rm c}(t)+\hat{n}_{\rm d}(t)=
\alpha^2+\frac{1}{2}\alpha\left(
\delta\hat{X}_{\rm a,0}(t)+ \delta \hat{X}_{\rm v,0}(t) +
\delta \hat{X}_{\rm a,0}(t-\tau)-
\delta \hat{X}_{\rm v,0}(t-\tau)\right),\\
\hat{n}_{\rm c}(t)-\hat{n}_{\rm d}(t)=
\alpha^2 \cos{\varphi} +
\frac{1}{2}\alpha\left(
\delta\hat{X}_{\rm a,-\varphi}(t-\tau)-
\delta\hat{X}_{\rm v,-\varphi}(t-\tau)+
\delta\hat{X}_{\rm a,\varphi}(t)+
\delta\hat{X}_{\rm v,\varphi}(t)\right).
\end{eqnarray}
where the quadrature component $\delta \hat{X}_{\rm a,\varphi}$ is defined by
$\delta \hat{X}_{\rm a,\varphi}={\rm e}^{i \varphi}\delta \hat{a}^{\dagger} +
{\rm e}^{-i \varphi}\delta \hat{a}$.

Then, via Fourier transform, the spectral components of the rf--fluctuations at
sideband frequency $\Omega$ are obtained. Since the phase shift of the spectral
components is given by $\Omega\tau=\theta$ (note that the Fourier transformation
${\cal F}$ gives ${\cal F} f(t-\tau)=e^{-i\Omega\tau} {\cal F}f(t)$), and the
optical phase is adjusted to $\varphi=\pi/2+2k\pi$ ($k$ is an integer), the
fluctuations of the sum and the difference photocurrents in frequency space read
as \begin{eqnarray} \delta\hat{n}_{\rm c}^{\Omega}+\delta\hat{n}_{\rm
d}^{\Omega}&=& \frac{1}{2}\alpha\left( \delta \hat{X}_{\rm a,0}^{\Omega}+{\rm
e}^{-i\theta}\delta \hat{X}_{\rm a,0}^{\Omega}+ \delta \hat{X}_{\rm
v,0}^{\Omega}-{\rm e}^{-i\theta} \delta \hat{X}_{\rm v,0}^{\Omega} \right), \\
\delta\hat{n}_{\rm c}^{\Omega}-\delta\hat{n}_{\rm d}^{\Omega}&=&
\frac{1}{2}\alpha\left( \delta\hat{X}_{a,\pi/2}^{\Omega}+
{\rm e}^{-i\theta}\delta\hat{X}_{\rm a,-\pi / 2}^{\Omega}
+\delta\hat{X}_{\rm v,\pi/2}^{\Omega}-
{\rm e}^{-i\theta}\delta\hat{X}_{\rm v,-\pi/2}^{\Omega}\right).
\label{gl6}
\end{eqnarray}
For $\theta=\pi$, the sum signal yields $\delta\hat{n}_{\rm
c}^{\Omega}+\delta\hat{n}_{\rm d}^{\Omega}=\alpha\delta \hat{X}_{\rm
v,0}^{\Omega}$ and the difference signal $\delta\hat{n}_{\rm
c}^{\Omega}-\delta\hat{n}_{\rm d}^{\Omega}=\alpha\delta \hat{X}_{\rm
a,\pi/2}^{\Omega}$ which is proportional to the spectral component $\Omega$ of
the phase quadrature of the initial field. The delay must therefore be chosen
such that a phase shift of $\pi$ between the two arms of the
interferometer is introduced at measurement frequency $\Omega_{\rm m}=2\pi
f_{\rm m}$ . Corresponding delay $\Delta L$ is then given by $\Delta
L=cT/2=c\pi/\Omega_{\rm m}=c/2 f_{\rm m}$ where c is the speed of light and T is
the period of the rf--signal at the frequency $\Omega_{\rm m}$.

\begin{figure}[h!]
\includegraphics{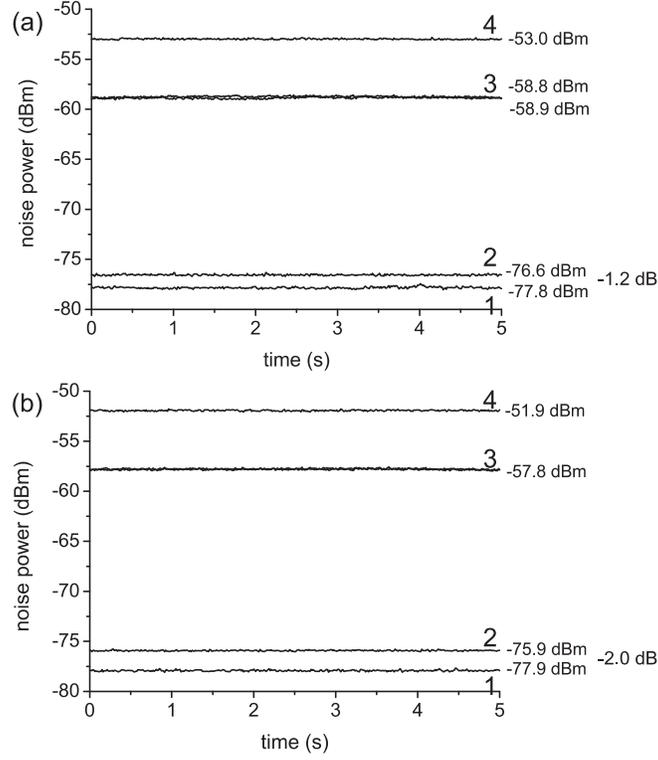}
\caption{\label{ergebnis}Correlations of the phase quadrature (a) and
the amplitude quadrature (b). In each graph the noise level (traces 1) of the
correlation signal is shown together with the corresponding shot noise level
(traces 2), the noise level of the individual beams (traces 3) and the signal
with the anticorrelations (traces 4). The traces were corrected by subtracting
the electronic noise, that was at $-84.7$dBm.}
\end{figure}

The experimental setup for the phase measuring interferometer is depicted
in Fig.~\ref{bild2}. It contains a Mach--Zehnder interferometer followed by a
balanced detection system using high efficiency InGaAs--photodiodes (Epitaxx
ETX 500). The first beam splitter in the interferometer is made
of a polarizing beam splitter and a $\lambda/2$--plate. It is therefore possible
to switch between phase measurement creating equal intensity in both arms and
amplitude measurement, where all light propagates through one arm.
The latter situation is equivalent to a balanced detection scheme,
in which the sum and the difference signal provides the amplitude noise and
the shot noise level respectively. The variances of the photocurrent
fluctuations were recorded with a pair of spectrum analyzers (8590L from HP) at
a resolution bandwidth of 300kHz and a video bandwidth of 30Hz. The measurement
time for each recorded trace as in Fig.\,\ref{ergebnis} was 5s. The difference
of the DC--powers of the two detectors served as error signal and was fed back
onto a piezo--mirror in the interferometer to stabilize the optical phase.

The quantum source that we characterize with our phase measuring device produces
intense quadrature entangled light pulses. An OPO pumped by a mode locked
Ti:Sapphire laser is used as light source. It produces pulses of 100fs at a
center wavelength of 1530nm and at a repetition rate of 82MHz. The nonlinear
Kerr effect experienced by intense pulses in optical fibers \cite{ROS91} is used
to generate amplitude squeezed light pulses employing an asymmetric fibre Sagnac
interferometer \cite{KRY98,SCHM98}. By use of the linear interference
of amplitude squeezed beams, intense entangled light was generated utilizing the
fibre optical setup described by Silberhorn et al.\cite{SIL01} To demonstrate
sub--shot noise performance of the interferometer each of the entangled beams
was directed into a phase measuring interferometer, we verified the correlations
of the detected photocurrents and compared them with the corresponding
shot noise level.

Working with a pulsed system, phase measurements can only be performed at
certain frequencies where interference occurs, as possible delays are governed
by the repetition frequency $f_{\rm rep}$ of the laser source $ \Delta
L=cnT_{\rm rep}=c n/f_{\rm rep}$ ($n$ is an integer number; $T_{\rm rep}$ is the
time between two pulses). Possible measurement frequencies are then given by
$f_{\rm m}=f_{\rm rep}/2n$. In our case, with a repetition rate of 82MHz, the
arm length difference must be a multiple of 3.66m, corresponding to the distance
between two successive pulses. For a frequency of 20.5MHz an arm length
difference of 7.32m is required. To achieve high interference contrast and hence
high efficiency of the interferometer, not only the phase fronts of the light
from the long and the short arm have to be matched carefully, but also the
temporal overlap of the pulses.

To characterize the entanglement source, first the squeezing resources used to
generate entanglement were investigated using the two interferometers in the
amplitude quadrature settings. In the experiment, $2.1$dB and $2.4$dB of
amplitude squeezing were observed for the two input beams respectively. The
level of squeezing is limited by the losses of the beams in the interferometers,
by the use of four imperfect detectors, and non-optimum balancing.

In a next step, the expected quantum correlations (anticorrelations) for the
phase (amplitude) quadrature were verified. These were checked by looking at the
noise level of the difference (sum) signal of each of the photocurrents of the
two entangled beams for the phase (amplitude) measurement. The phase
correlations are $-1.2$dB below the shot noise level (see trace 1 in Fig.
\ref{ergebnis}a) corresponding to a squeezing variance $\langle(\delta
\hat{X}_{1,\pi/2}^{\Omega}-\delta \hat{X}_{2,\pi/2}^{\Omega})^2
\rangle/2=0.76\pm0.02$ while the amplitude correlations were at $-2.0$dB (see
trace 1 in Fig. \ref{ergebnis}b) corresponding to a squeezing variance
$\langle(\delta \hat{X}_{1,0}^{\Omega}+\delta \hat{X}_{2,0}^{\Omega})^2
\rangle/2=0.63\pm0.02$ (indices 1,2 refer to two entangled modes). The
discrepancy between the noise level of the amplitude and the phase quadrature
measurement comes from imperfect mode matching in the latter case. We achieved a
visibility of 85\%, introducing additional losses of 28\% for the phase
measurement. Thus, instead of $-2$dB of correlations as in the amplitude
measurement we do not expect correlations stronger than $-1.3$dB in the phase
measurement, which agrees with the measurement results.

The noise level of the phase quadrature measurement of the individual entangled
beams (traces 3 in Fig.\ref{ergebnis}a) is $6$dB below the corresponding
sum--signal of these two beams (trace 4), which is due to the quantum
correlations of the entangled beams. The same applies for the noise levels of
the amplitude quadrature. Note that the noise level of the individual beams is
$\sim\!\!18$dB above the shot noise level due to the high degree of excess
phase noise of the initial squeezed light, introduced by self phase modulation
as well as by guided acoustic wave Brillouin scattering \cite{SHE85a}.

In this experiment, for the first time, non--classical correlations in the
phase quadrature have been observed directly for intense pulsed light. No
additional interaction between the individual entangled beams \cite{SIL01} is
necessary to demonstrate sub shot noise correlations in the phase quadrature.
According to the non--separability criterion by Duan \cite{DUA00} et al. and Simon
\cite{SIM00} for Gaussian systems, the existence of nonclassical
correlations implies $\Delta=\sqrt{\langle(\delta \hat{X}_{1,\pi/2}^{\Omega}-\delta
\hat{X}_{2,\pi/2}^{\Omega})^2 \rangle \langle(\delta
\hat{X}_{1,0}^{\Omega}+\delta \hat{X}_{2,0}^{\Omega})^2 \rangle/4}<1$. Since
$\Delta=\sqrt{0.76\cdot 0.63}<1$ we can conclude qualitatively that the state is
non--separable. Because our bipartite Gaussian state is symmetric we can also
quantify the entanglement of formation $E_F$ using proposition 2 in ref.
\cite{GIE03} and we find $E_F=0.22\pm0.02$.

The implemented phase measurement device opens the possibility of performing a
variety of experiments in the field of quantum information and communication,
where intense light beams are used and phase measurements are required. For
example, a quantum key distribution scheme relying on quadrature
entanglement\cite{SIL02} could be implemented without sending a local oscillator
together with the signal beams. Also the full experimental demonstration of
continuous variable entanglement swapping using intense light beams seems to be
possible\cite{GLOE03}.

In conclusion, we experimentally demonstrated a setup for measuring the
quantum fluctuations of the phase quadrature on intense, pulsed light without
using a separate local oscillator or a resonator. We prove sub--shot noise
resolution by resolving phase correlations below the shot noise level on a pair
of quadrature entangled beams.

We would like to thank N. L\"utkenhaus, P.\,Grangier and J.\,Trautner for useful
discussions and A.\,Berger for building the electronic circuits for the phase
lock. This work has been supported by  the Schwerpunktprogramm 1078 of the
Deutsche Forschungsgemeinschaft and the Kompetenznetzwerk
Quanteninformationsverarbeitung der Bayerischen Staatsregierung (A8). U. L.
Andersen gratefully acknowledges financial support of the Alexander von Humboldt
foundation.

*Present address, Clarendon Laboratory, University of Oxford, Parks Road, Oxford
OX1 3PU, UK

%
%

\newpage

\newpage

\end{document}